\documentstyle[12pt,epsf]{article}
\begin{document}
\renewcommand{\thefootnote}{\fnsymbol{footnote}}
\begin{flushright}
KEK-preprint-94-106\\
DPNU-94-38\\
NWU-HEP 94-04\\
TIT-HPE-94-09\\
TUAT-HEP 94-04\\
OCU-HEP 94-06\\
PU-94-686\\
INS-REP 1051\\
KOBE-HEP 94-05\\
\end{flushright}
\vskip -3cm
\epsfysize3cm
\epsfbox{kekm.epsf}
\begin{center}
{\large \bf
$K^0(\overline{K^0})$ Production in Two-Photon Processes at TRISTAN
\footnote{
to be published in Phys. Lett. {\bf B}.
}
}\\
\vskip 0.5cm
(TOPAZ Collaboration)\\
\vskip 0.5cm
\underline{R.Enomoto}$^a$\footnote{Internet address: enomoto@kekvax.kek.jp.},
K.Abe$^b$, T.Abe$^b$, I.Adachi$^a$,
K.Adachi$^c$, M.Aoki$^d$, M.Aoki$^b$, S.Awa$^c$,
K.Emi$^e$, H.Fujii$^a$, K.Fujii$^a$,T.Fujii$^f$, J.Fujimoto$^a$,
K.Fujita$^g$, N.Fujiwara$^c$, H.Hayashii$^c$,
B.Howell$^h$, N.Iida$^a$, R.Itoh$^a$, Y.Inoue$^g$, H.Iwasaki$^a$,
M.Iwasaki$^c$, K.Kaneyuki$^d$, R.Kajikawa$^b$,
S.Kato$^i$, S.Kawabata$^a$, H.Kichimi$^a$, M.Kobayashi$^a$,
D.Koltick$^h$, I.Levine$^h$, S.Minami$^d$,
K.Miyabayashi$^b$, A.Miyamoto$^a$, K.Muramatsu$^c$, K.Nagai$^j$,
K.Nakabayashi$^b$, E.Nakano$^b$,
O.Nitoh$^e$, S.Noguchi$^c$, A.Ochi$^d$, F.Ochiai$^k$,
N.Ohishi$^b$, Y.Ohnishi$^b$, Y.Ohshima$^d$,
H.Okuno$^i$, T.Okusawa$^g$,T.Shinohara$^e$, A.Sugiyama$^b$,
S.Suzuki$^b$, S.Suzuki$^d$, K.Takahashi$^e$, T.Takahashi$^g$,
T.Tanimori$^d$, T.Tauchi$^a$, Y.Teramoto$^g$, N.Toomi$^c$,
T.Tsukamoto$^a$, O.Tsumura$^e$, S.Uno$^a$, T.Watanabe$^d$,
Y.Watanabe$^d$, A.Yamaguchi$^c$, A.Yamamoto$^a$,and M.Yamauchi$^a$\\
\vskip 0.5cm
{\it
$^a$KEK, National Laboratory for High Energy Physics, Ibaraki-ken 305, Japan \\
$^b$Department of Physics, Nagoya University, Nagoya 464, Japan\\
$^c$Department of Physics, Nara Women's University, Nara 630, Japan \\
$^d$Department of Physics, Tokyo Institute of Technology, Tokyo 152, Japan\\
$^e$Dept. of Appl. Phys., Tokyo Univ. of Agriculture and
Technology, Tokyo 184, Japan\\
$^f$Department of Physics, University of Tokyo, Tokyo 113, Japan\\
$^g$Department of Physics, Osaka City University, Osaka 558, Japan \\
$^h$Department of Physics, Purdue University, West Lafayette, IN 47907, USA \\
$^i$Institute for Nuclear Study, University of Tokyo, Tanashi,
Tokyo 188, Japan \\
$^j$The Graduate School of Science and Technology, Kobe University, Kobe 657,
Japan \\
$^k$Faculty of Liberal Arts, Tezukayama University, Nara 631, Japan \\
}
\end{center}
\newpage
\begin{abstract}
\baselineskip 20pt
We have carried out an inclusive measurement of $K^0(\overline{K^0})$
production in two-photon
processes at TRISTAN. The mean $\sqrt{s}$ was 58 GeV and the
integrated luminosity was 199 pb$^{-1}$.
High-statistics $K_s$ samples were obtained under such conditions as
no-, anti-electron, and remnant-jet tags.
The remnant-jet tag,
in particular, allowed us, for the first time, to measure
the cross sections
separately for the resolved-photon and direct processes.
\end{abstract}
\newpage
\baselineskip 24pt
\section{Introduction}
Excess over theoretical predictions was
reported in references \cite{exclusive,inclusive}, concerning
charm pair production in two-photon processes.
These references extensively discussed
possibilities to explain the excess by
increasing the predicted cross sections of two-photon processes.
In order to help sorting out these possibilities,
we have carried out an inclusive measurement of $K^0(\overline{K^0})$
mesons in two-photon
processes.
With the point-like process (direct process \cite{direct}) where
$e^+e^-\rightarrow e^+e^-s\bar{s}$ is strongly suppressed
due to the small value of $Q^4_s$ ($Q_s$ is s quark's charge),
$K_s$'s, particularly in the high-$P_T$ region, come mainly from
$c\bar{c}$ production \cite{mark2}.
The use of $K_s$'s also enhances the sensitivity to
two-photon processes at low $P_T$.

There are additional
advantages in this $K_s$ analysis,
which are high statistics and low systematics,
thanks to high acceptance and use of only a central tracking device,
respectively.
Moreover, using the low angle calorimeter \cite{fcl}, we could
obtain the cross section for the resolved photon process
\cite{resolved} separately.
We can thus test some of the possibilities suggested
in reference \cite{inclusive}: we proposed to
use a low charm quark mass (1.3 GeV),
the next-to-leading-order correction (NLO), and the
intrinsic parton $P_T$ inside photon in order to
explain an excess observed in
the $c\bar{c}$ cross section especially in the high-$P_T$ region.

Notice also that this analysis is hardly affected by
$\tilde{t}$ (superpartner of
the top quark) pair production \cite{stop}
discussed in references \cite{exclusive,inclusive},
even if present, since
the high-$Q^2$ decay of $c\rightarrow \overline{K^0} X$ smears
the $P_T$ distribution
of the $K_s$'s and thus diminishes
the sensitivity to the $\tilde{t}$
pair production.

\section{Event selection}
The data used in this analysis were obtained with the TOPAZ detector
at the TRISTAN $e^+e^-$ collider, KEK\cite{topaz,tpc}.
The mean $\sqrt{s}$ was 58 GeV and the integrated luminosity
was 199 pb$^{-1}$.
A forward calorimeter (FCL) ,which covered
$0.98<|\cos\theta|<0.998$ ($\theta$ is the
polar angle, i.e., the angle with respect to the electron beam),
was installed in the course of the experiment.
The FCL was made of bismuth germanate
crystals (BGO), and was used to anti-tag the beam electrons (positrons)
and to tag hadrons (remnant-jets) \cite{fcl} for this study.
We could thus select collisions of almost-real photons
including resolved
photon processes\cite{resolved}.
The integrated luminosity of the data with the FCL
detector was 175 pb$^{-1}$.

A description of our trigger system can be found in
reference \cite{trigger}. The requirement for the charged track trigger
was two or more tracks with an opening angle of greater than
45-90 degrees. The $P_T$ threshold for charged particles was 0.3-0.7 GeV,
varied depending on beam conditions.

The event selection criteria were as follows:
three or more charged particles ($P_T>0.15$ GeV, $|\cos\theta|<$0.77),
the invariant mass ($W_{VIS}$) of visible particles
($|\cos\theta|<$0.77) had to be greater than 2 GeV,
the event-vertex position had to be consistent with the interaction point,
and the visible energy
had to be less than 25 GeV.
In total, 220378 events were selected.

\section{Monte-Carlo simulation}
In order to estimate the acceptances and backgrounds in this analysis,
we used the following Monte-Carlo simulation programs.
For the generation of single-photon-exchange hadronic events, we used
JETSET6.3\cite{lund} with the parameter values given in
reference \cite{adachi}. Details concerning the
event generation of
direct
as well as
resolved-photon
and
vector meson dominance (VDM) processes
can be found in references \cite{exclusive,inclusive,hayashii}.
Here, we just note the following points.
For $c\bar{c}$ generation,
we used the current charm quark mass of 1.3 GeV
to calculate cross sections for point-like processes and a
constituent charm quark mass of 1.6 GeV
for hadronization procedure, and made the
next-to-leading order (NLO) correction,
whose details can be found in
references \cite{exclusive,inclusive,drees}.
Light-quark generation was carried out by using the lowest
order (LO) formula with a $P_T^{min}$ cut of 2.5 GeV.
We used the parton density functions by Levy-Abramowicz-Charchula set-1 (LAC1)
\cite{lac1} for the resolved-photon process.
Generated events were processed through the standard TOPAZ detector
simulation program \cite{adachi}, in which
hadron showers were simulated with an extended version of
GHEISHA 7 \cite{gheisha}.
Its data on hadron
interactions with nuclei had been updated to fit various
experimental cross sections.

Using the above-mentioned Monte-Carlo simulations, the trigger
efficiency for the sum of the direct and resolved photon processes was
estimated to be 79\%, 97\% of which represented charged trigger events.
The event-selection efficiency after the trigger was obtained to
be 80\%.

\section{Tagging conditions}
The tagging conditions were as follows. For
anti-electron tagging, there had to be no
energy deposit of more than 0.4$E_b$
in $|\cos\theta|<0.998$ (anti-electron tag or anti-tag),
where $E_b$ is the
beam energy.
This selected events from collisions of almost
real photons, for which the accuracy of the equivalent photon approximation
was expected to be reliable at the 1\% level.
When the energy cut was lowered, mis-anti-tag due to
beam remnant hadrons (remnant-jets) became significant,
as predicted by the Monte-Carlo simulations.
The energy distribution of the maximum-energy cluster in FCL
is shown in Figure \ref{fclhad}.
The horizontal scale is normalized at the beam energy.
The Monte-Carlo predictions for single-photon-exchange,
VDM, and resolved-photon processes are shown by histograms.
The peak around 1 was explained by the direct processes.
This implies a possibility to tag the resolved photon process
by requiring, for instance, 500 MeV $<E_{FCL}<$ 0.25$E_b$,
where $E_{FCL}$ is the energy deposit in the FCL (remnant-jet tag
or rem-tag).
We did not use any hadron shape information because of large
segmentation and lack of tracking information.
The yield of the remnant-jet tag events agreed with our Monte-Carlo
simulation within 5\% level.
Our analysis is the first trial that uses this tag.
An event selection without these two tags is called, hereafter,
``no-electron tag" or ``no-tag".
The fractions of electron and remnant-jet tag events to no-tag ones
were obtained to be 2.4 and 47\% of the selected events, respectively.

In the Monte-Carlo simulations, we used the equivalent photon approximation
with the
photon flux expression given in reference \cite{flux}.
We set the $Q^2_{\gamma}$ limit at the smaller of $P_{T,q}^2+m_q^2$
and the anti-tag limit
[$2E_b^2(1-x_{\gamma})(1-\cos\theta_{max}):~x_{\gamma}=0.4,~\theta_{max}
=3.2$ degrees],
where $P_{T,q}$ and $m_q$ are
the transverse momentum and the mass
of a quark, respectively \cite{whit}.

The tagging efficiency of the remnant-jet tag for the resolved photon
process was estimated to be 72\%
without assuming FCL noise which will be described later.
We tried two ways of generating remnant partons:
one along the beam direction,
and the other using a Gaussian distribution of $P_T$-width 0.44 GeV
with respect to the
beam axis. Two methods differed in acceptance only by 3\%.
On the other hand, the tagging efficiency of the remnant-jet tag
for the direct process was estimated to be 10.8\%
without assuming FCL noise.


\section{Analysis}
The charged-track selection criteria for the $K_s$ analysis were as follows:
for each track
$P_T$ had to be greater than 0.15 GeV,
dE/dx had to be consistent with the pion
assumption ($\chi^2_{\pi^{\pm}}<10$),
$|\cos\theta|$ had to be less than 0.77, and the closest
approach to the interaction point in the XY-plane (perpendicular
to the beam axis) had to be greater than 1 cm.
Using these selected tracks,
we looked for opposite-sign pairs with an opening angle
less than 90 degrees,
and carried out secondary vertex reconstructions
three-dimensionally.
Finally, we demanded these pairs
to be consistent with the assumption that they came
from the event vertices with a flight length larger than 3 cm.
The invariant-mass distributions of these candidate
$\pi^+\pi^-$ pairs are plotted in Figure \ref{mass}
for the three tagging conditions, respectively.
These invariant-mass spectra were fitted with the sum of a second-order
polynomial and a Gaussian distribution
and the peak entries were obtained for no-,
remnant-jet, and electron tags
to be $893\pm34$, $364\pm22$, and $75.8\pm9.5$ $K_s$'s,
respectively.
The peak position and the width were consistent with the detector
simulation.
In order to derive the differential $P_T$ cross sections, we divided
$P_T$ into ten bins, as shown in Table \ref{tcross}.
Notice that
even the lowest-statistics bin gave a 5.1-$\sigma$ $K_s$ peak
(in the no-electron tag case).

\section{Background subtractions}
Single-photon-exchange process was a largest background
especially for high-$P_T$ $K_s$.
This can be reduced when a cut was applied on the total visible energy.
We, however, did not carry out this, because we did not want
to reduce the acceptance for high-$P_T$ $K_s$.
The contamination from the single-photon-exchange process was
estimated and subtracted using the Monte-Carlo
simulation, on a bin by bin basis.
The background fractions
for no-tag were 6.0, 6.1, 6.5, 8.1, 10.0, 11.0, 14.3, 17.3,
43.4, and 48.1\%, respectively,
for the $P_T$ bins shown in Table \ref{tcross},
being strongly $P_T$ dependent.
We estimated these fractions for anti- and remnant-jet tags. They were
consistent with the above values within statistical errors.
The background from beam gas interactions was estimated using the
off-vertex events in the beam direction:
there was a vacuum leak in the beam pipe for some period.
The beam gas
contribution
for no-tag was 7.8\% on the average, and was subtracted from the data.
For anti- and remnant-jet tags, the above value became slightly
large (9.8\%), i.e., electron-tag sample was free from the beam-gas
background.
FCL noise hits were studied by analysing random trigger
and Bhabha events.
The probability of noise hits with $E_{FCL}>$0.5 GeV
was estimated to be 12.1\%,
while for hits with $E_{FCL}>0.4E_b$, it reduces to 0.1\%.
The FCL noise was also related to the vacuum leak.
In the Monte-Carlo simulations, we added noise hits randomly
in accordance with the observed noise hit
probability in order to reliably estimate the tagging efficiencies.
Using this, the tagging efficiencies of
the remnant-jet tag for the resolved-photon and direct
processes were estimated to be 75\% and 21\%, respectively.

\section{Systematic errors}
The systematic errors were estimated, bin by bin, as follows.
For the trigger, we added some extra noise hits in the tracking chambers
in the simulations.
For the event selection and the $K_s$ reconstruction,
we changed the cut values and evaluated
systematic errors as
the cross-section differences.
We also changed the pulse-height threshold in the TPC simulation
to evaluate the effects on its tracking efficiency.
We added the obtained systematic errors quadratically, on a bin by bin
basis.
The average over the $P_T$ bins of the systematic errors was 12\%,
of which
the cut dependence in the event selection
was the dominant source.
These systematic errors were quadratically added to
statistical errors.
We also checked the acceptance ambiguity due to the parametrization
dependence
of the resolved-photon processes,
by comparing the LAC1 \cite{lac1} and
Drees-Grassie [DG] \cite{dg} parametrizations.
The acceptance difference was estimated to be 5.9\%, which is small
compared to the systematic errors shown above.

\section{Results}
The $P_T$ differential cross sections were obtained
from the number of reconstructed $K_s$'s in each bin and
its corresponding
efficiency estimated with the Monte-Carlo simulations described
previously.
They are listed in Table \ref{tcross} and plotted in Figures
\ref{fcross} (a) - (e) for the three tagging conditions
and two subtraction schemes,
respectively. Figure \ref{fcross} (a) is for anti-electron tag
events.
In the remnant-jet tag events, the Monte-Carlo simulation predicted
a significant contamination from VDM events.
The tagging efficiency for the VDM process was estimated to
be $\sim$61\%, slightly smaller than that of the resolved-photon
process. In addition there was a large ambiguity in the cross
section of the VDM process. Therefore we calculated the cross
sections using two subtraction schemes. Figures \ref{fcross}
(b) and (e) were obtained by subtracting the VDM contribution
predicted by the Monte-Carlo simulation (VDM subtraction) for
remnant-jet and anti-remnant-jet tags, respectively.
Here, the ``anti-remnant-jet tag" cross section was
obtained by subtracting the remnant-jet tag cross section from
that of the anti-electron tag.
Figures \ref{fcross} (c) and (d) were obtained without the
VDM subtraction.
In Figures \ref{fcross} (b) and (d), the contribution
of the direct process was subtracted, since the
uncertainty in the prediction of this process was
considered to be small.
The histograms in Figures \ref{fcross} (a) - (e) are the
Monte-Carlo predictions:
the cross-hatched, singly-hatched, and open areas are
predictions for the direct,
resolved-photon (LAC1), and VDM processes.

\section{Discussions}
The fraction of charm events was studied using the
above-mentioned Monte-Carlo simulations.
We found 55\% of
these events with $P_T(K_s)>$ 1.4 GeV were of charm origin.
On the other hand,
only 30\% of the
events with charged tracks of
$P_T>$ 1.4 GeV were from $c\bar{c}$ pairs.
Also, the Monte-Carlo simulations predicted that 70\% of these
high-$P_T$ charm events originated from the direct process.
In this study, we derived six types of cross sections using
different tagging conditions and subtraction schemes.
We can therefore separately compare
each cross section with the theoretical prediction for each
process.

Firstly, about 30\% of the high-$P_T$ events can be explained as the
electron-tagged events (see Table \ref{tcross}).
Secondly, for the anti-tag cross section [Figure \ref{fcross} (a)],
the agreement, especially in the lower-$P_T$ region, is reasonably good,
considering the ambiguities due to the VDM process.
They are, however, higher than the theoretical
predictions in the high-$P_T$
region ($P_T>1.2$ GeV)
 by 2.2$\sigma$.

The cross section with the anti-remnant-jet tag is consistent with the
predictions of the direct and VDM processes [Figures \ref{fcross} (c)
and (e)].
The hypothetical $\tilde{t}$ pair production with
$m_{\tilde{t}}$=15 and $m_{\tilde{\gamma}}$=12.7 GeV
\cite{exclusive,inclusive} is expected to increase
these cross sections by
1.7, 1.1, and 0.3 pb/GeV in the highest three $P_T$ bins,
respectively, which are smaller than the experimental errors.
Therefore we could not discuss this hypothesis by this analysis result.
Note also that this hypothesis turned out to be
inconsistent with the recent
search by the VENUS collaboration \cite{shirai}.

The dominant source of
the discrepancy lies in Figures \ref{fcross} (b) and (d),
i.e., in the remnant-jet tag sample
(we considered that this sample
was dominated by the resolved-photon and/or
VDM processes):
the spectrum looks harder than the prediction of the
resolved photon process.
The histograms in Figures \ref{fcross} (b) and (d) have already been
corrected by the
NLO factor (0.5$P_{T,c}$+0.54; $P_{T,c}$ is a $P_T$ of a charm quark)
\cite{exclusive,inclusive,drees}.
This factor, which is due to the presence of off-shell gluon emissions
from resolved quarks, is large especially in the high $P_T$ region.
The measured spectrum is harder than
the LO prediction,
and is close to the spectrum of the
direct process.
Our data suggest the importance of the NLO correction. The NLO
correction to the light quark events, which is absent from our
present Monte-Carlo generator,
is also considered to be necessary.
In addition, the intrinsic parton $P_T$ (we used 0.44 GeV
with a Gaussian distribution) is necessary.
In summary, the theoretical prediction agrees with our
data to a reasonably good accuracy, justifying our
parametrization of the Monte-Carlo generation.

In order to check if the parton density functions
have anything to do with the discrepancy,
we compared our remnant-jet-tag data [Figure \ref{fcross} (d)]
with the predictions from six sets of parametrizations by
Hagiwara, Tanaka, Watanabe, and Izubuchi [WHIT1-6] \cite{whit}.
A systematic analysis on gluon distributions can be carried out
using these parametrizations.
The results are shown in Figures \ref{fres} (a) and (b),
where the histograms are the predictions by the WHIT 1-6 parametrizations
with $P_T^{min}$'s of 2.0 and 2.5 GeV.
Notice that the predictions are rather sensitive to the $P_T^{min}$ cut
and the VDM contribution,
especially in the low-$P_T$ region; $P_T<$ 1 GeV.
There are some possible combinations which reproduce the high-$P_T$
cross sections well. It is, however, necessary to improve the
prediction of the VDM process.

\section{Conclusion}
We carried out an inclusive measurement of $K^0(\overline{K^0})$
in two-photon
processes at TRISTAN. The mean $\sqrt{s}$ was 58 GeV and the
integrated luminosity was 199 pb$^{-1}$.
High-statistics $K_s$ samples were obtained under such conditions as
no-, anti-electron, and remnant-jet tags.
Especially with the remnant-jet tagging,
we could unambiguously extract the contribution
from the resolved photon process.
Comparisons with the theoretical predictions were carried out.
Our results agreed with the theoretical predictions with a
low charm mass ($m_c$=1.3 GeV),
intrinsic parton $P_T$ inside photon
($\sigma_{P_T}$=0.44 GeV),
and the NLO corrections, i.e., those obtained by the
previous results of the $D^{*\pm}$ analysis \cite{exclusive,inclusive}.

\section*{Acknowledgement}
We would like to thank Prof. H. Terazawa (INS) for discussions
concerning the
theoretical predictions. We also thank Drs. K. Hagiwara, M. Tanaka,
I. Watanabe (KEK), and Mr. T. Izubuchi (Univ. of Tokyo) for instructions
on the WHIT parametrizations.

\newpage
\section*{Table \ref{tcross}, R. Enomoto et al., Physics Letters B.}
\begin{table}[h]
\begin{tabular}{ccccccc}
\hline
\hline
tag cond. & no-tag & anti-tag & rem-tag & anti-rem-tag
& rem-tag & anti-rem-tag\\
VDM subt. & - & - & yes & no & no & yes \\
\hline
$P_T$ range& \multicolumn{6}{c}{cross sections} \\
 (GeV) & \multicolumn{6}{c}{(pb/GeV)} \\
\hline
 0.3-0.55
 & 889$\pm$ 248& 810$\pm$ 247&  43$\pm$  52& 767$\pm$ 247
 & 613$\pm$ 237& 197$\pm$ 282\\
 0.55-0.65
 & 839$\pm$ 166& 783$\pm$ 168&  53$\pm$  42& 731$\pm$ 170
 & 520$\pm$ 161& 263$\pm$ 202\\
 0.65-0.75
 & 703$\pm$ 114& 593$\pm$ 109&  99$\pm$  34& 494$\pm$ 108
 & 556$\pm$ 117&  37$\pm$ 137\\
 0.75-0.9
 & 415$\pm$  73& 371$\pm$  70&  34$\pm$  20& 337$\pm$  69
 & 241$\pm$  62& 129$\pm$  74\\
 0.9-1.05
 & 236$\pm$  32& 213$\pm$  32&  41$\pm$  14& 172$\pm$  33
 & 161$\pm$  33&  52$\pm$  42\\
 1.05-1.2
 & 162$\pm$  24& 116$\pm$  22&  14$\pm$  10& 102$\pm$  23
 &  51$\pm$  20&  65$\pm$  28\\
 1.2-1.4
 &  70$\pm$  12&  62$\pm$  12&  16$\pm$   7&  46$\pm$  14
 &  30$\pm$  10&  32$\pm$  15\\
 1.4-1.7
 &37.7$\pm$ 8.3&30.6$\pm$ 7.6&15.8$\pm$ 5.3&14.8$\pm$ 8.0
 &19.1$\pm$ 6.1&11.5$\pm$ 8.3\\
 1.7-2.5
 & 7.2$\pm$ 2.2& 4.8$\pm$ 2.0& 4.7$\pm$ 1.8& 0.1$\pm$ 2.5
 & 4.8$\pm$ 1.9& 0.0$\pm$ 2.5\\
 2.5-5
 &1.78$\pm$0.73&1.63$\pm$0.70&0.98$\pm$0.64&0.65$\pm$0.93
 &0.98$\pm$0.64&0.65$\pm$0.93\\
\hline
\hline
\end{tabular}
\caption{
\baselineskip 20pt
Differential cross section of $K^0(\overline{K^0})$ versus $P_T$ (GeV),
$d\sigma/dP_T$ (pb/GeV), for $|\cos\theta|<0.77$. Six cases
are listed: no-tag , anti-electron tag, remnant-jet tag
with VDM subtraction,
anti-remnant-jet tag without VDM subtraction,
remnant-jet tag without VDM subtraction,
and anti-remnant-jet tag with VDM subtraction,
which are described in the text.
}
\label{tcross}
\end{table}
\newpage
\section*{Figure captions}
\begin{figure}[h]
\caption{
Distribution of the energy fractions (normalized at the beam energy)
of the maximum-energy clusters in FCL.
The points with error bars are experimental data.
The histograms are the prediction by the Monte-Carlo simulation;
the cross-hatched area is single-photon-exchange process,
the singly-hatched one is VDM, and the open one is resolved-photon
process.}
\label{fclhad}
\caption{
Invariant-mass spectra of $\pi^+\pi^-$'s. The solid
histogram is for the no-electron tag, the dashed one is for
the remnant-jet tag, and the dotted one is for the
beam-electron tag.
}
\label{mass}
\caption{
Differential cross section of $K^0(\overline{K^0})$ versus $P_T$ (GeV),
$d\sigma/dP_T$ (pb/GeV), for $|\cos\theta|<0.77$. Five cases
are plotted: (a) anti-electron tag, (b) remnant-jet tag
with the VDM and direct process subtraction,
(c) anti-remnant-jet tag without the VDM subtraction,
(d) remnant-jet tag without the VDM subtraction,
and (e) anti-remnant-jet tag with the VDM and direct process
subtraction, as
described in the text.
Processes which we expected to show are;
(b) resolved-photon, (c) direct and VDM, (d) VDM and resolved, and
(e) direct processes.
The histograms are the theoretical predictions which are described in
the text. The open area is for the VDM, the singly-hatched one is for
the resolved photon process, and the cross-hatched one is for
the direct process.
}
\label{fcross}
\caption{
Differential cross section of $K^0(\overline{K^0})$ versus $P_T$ (GeV),
$P_T^4d\sigma/dP_T$ (pb$\cdot$GeV),
for $|\cos\theta|<0.77$ for the remnant-jet
tag without the VDM subtraction.
The hatched areas are the predictions by the VDM Monte-Carlo simulation.
The histograms are predictions by the WHIT 1-6 parametrizations.
The solid one is WHIT1, the dashed one WHIT2, the dot-dashed
one  WHIT3,
the dotted one WHIT4, the scarce-dotted one WHIT5, and the
short-dashed one WHIT6.
Two values of $P_T^{min}$'s were used, i.e., (a) 2.0 and (b)
2.5 GeV.
}
\label{fres}
\end{figure}
\newpage
\section*{Figure \ref{fclhad}, R. Enomoto et al., Physics Letters B.}
\epsfysize10cm
\epsfbox{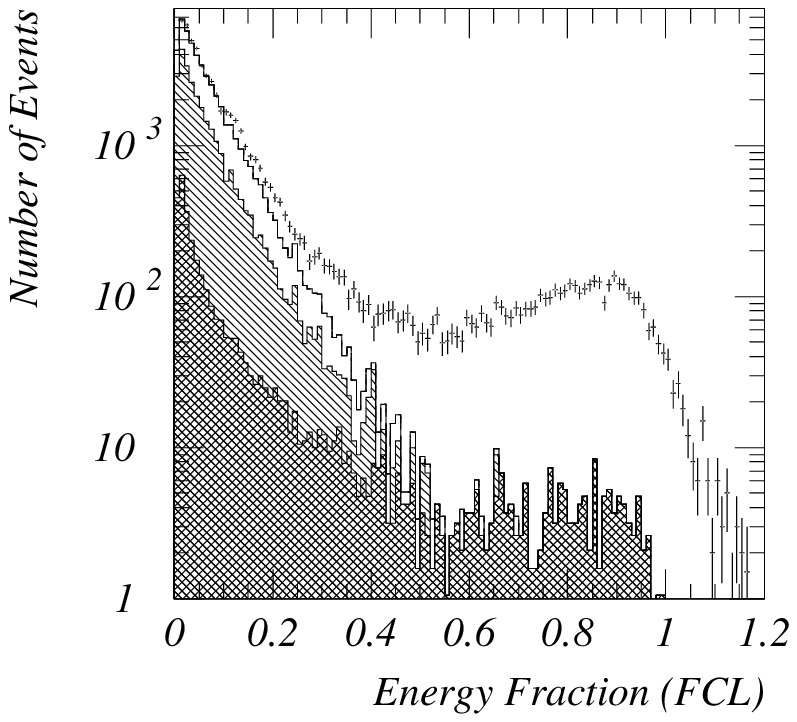}
\newpage
\section*{Figure \ref{mass}, R. Enomoto et al., Physics Letters B.}
\epsfysize10cm
\epsfbox{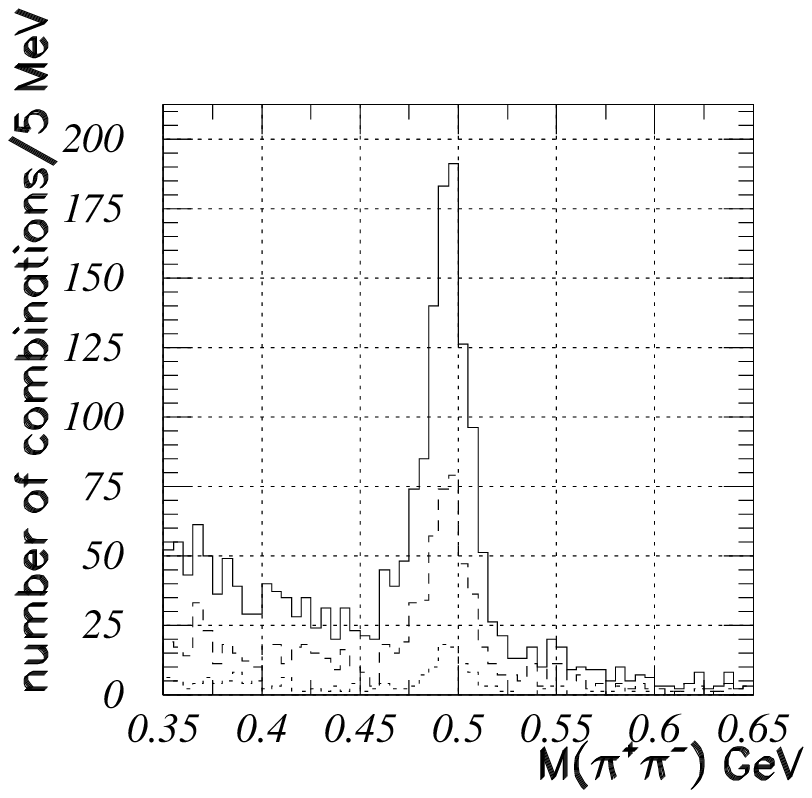}
\newpage
\section*{Figure \ref{fcross}, R. Enomoto et al., Physics Letters B.}
\epsfysize20cm
\epsfbox{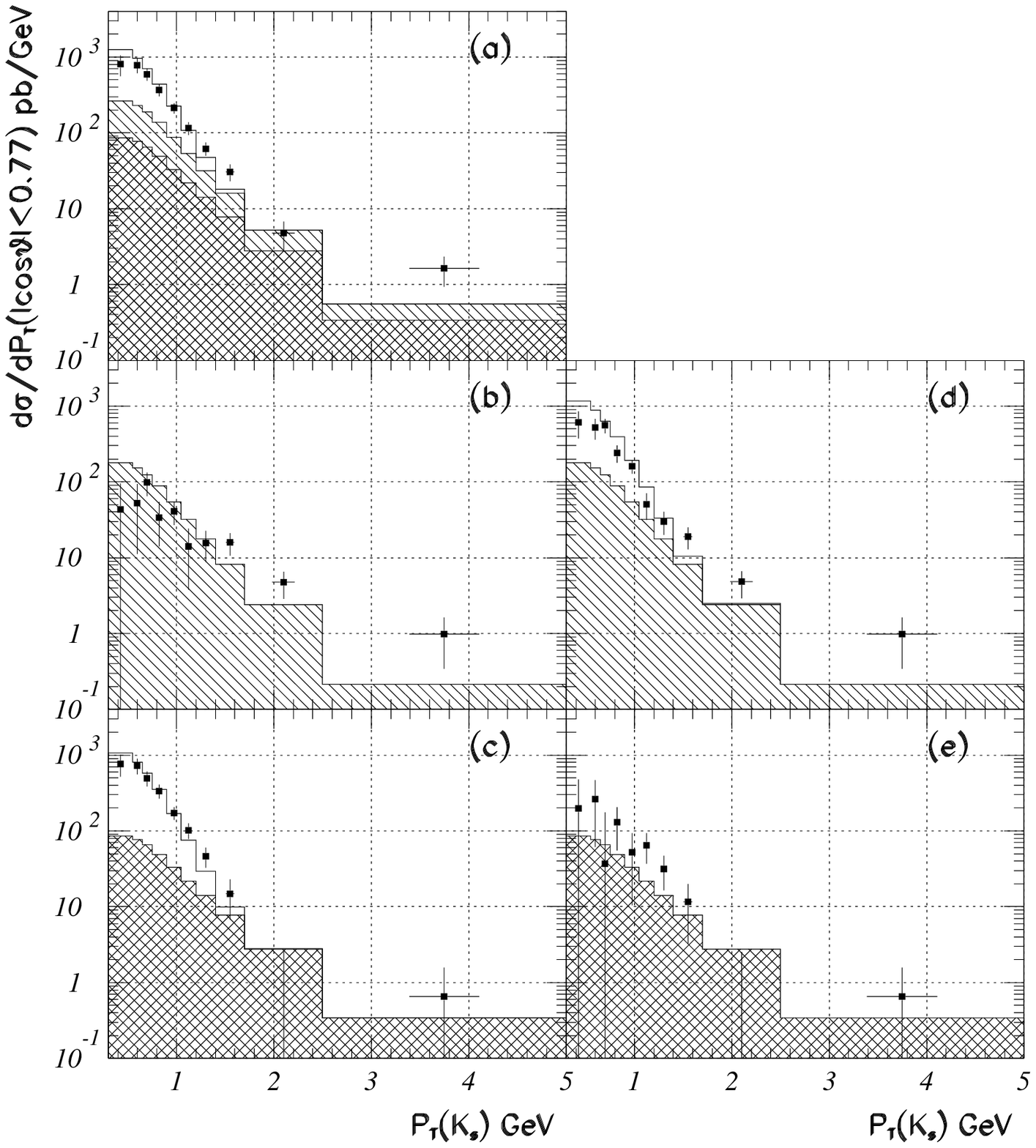}
\newpage
\section*{Figure \ref{fres}, R. Enomoto et al., Physics Letters B.}
\epsfysize20cm
\epsfbox{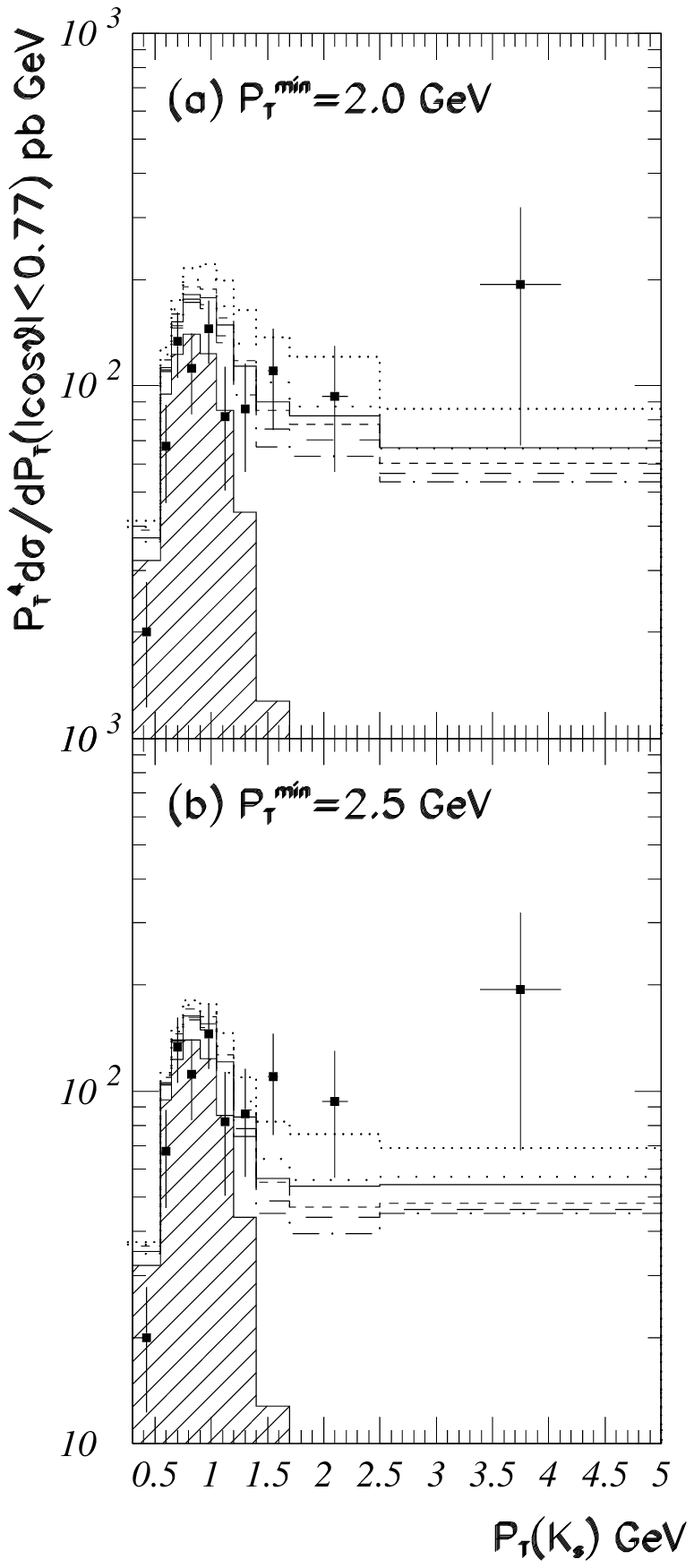}
\end{document}